\documentclass[graybox]{svmult}

\usepackage{type1cm}        
\usepackage{makeidx}         
\usepackage{graphicx}        
\usepackage{multicol}        
\usepackage[bottom]{footmisc}

\usepackage{newtxtext}       %
\usepackage{newtxmath}       

\RequirePackage{hyperref}
\RequirePackage{tabularx}
\RequirePackage{siunitx}
\RequirePackage{tikz}
\usetikzlibrary{shapes}
\definecolor{Orange}{cmyk}{0, 0.5, 1, 0}
\definecolor{SkyBlue}{cmyk}{0.8, 0, 0, 0}
\definecolor{BluishGreen}{cmyk}{0.97, 0, 0.75, 0}
\definecolor{Yellow}{cmyk}{0.1, 0.05, 0.9, 0}
\definecolor{Blue}{cmyk}{1, 0.5, 0, 0}
\definecolor{Vermillion}{cmyk}{0, 0.8, 1, 0}
\definecolor{ReddishPurple}{cmyk}{0.1, 0.7, 0, 0}
\definecolor{magenta1}{cmyk}{0,1,0,0}
\definecolor{grey}{RGB}{131,131,131}
\definecolor{colx}{cmyk}{1, 0.5, 0, 0}
\definecolor{coly}{cmyk}{0.97, 0, 0.75, 0}
\definecolor{colz}{cmyk}{0, 0.8, 1, 0}
\definecolor{colxy}{cmyk}{0, 0.5, 1, 0}

\newcommand{\grayline}{\textcolor{gray}{\protect\rule[0.5ex]{1.2em}{1pt}}}
\newcommand{\graydashed}{\textcolor{gray}{\protect\rule[0.5ex]{0.2em}{1pt}}\textcolor{white}{\protect\rule[0.5ex]{0.2em}{1pt}}\textcolor{gray}{\protect\rule[0.5ex]{0.2em}{1pt}}\textcolor{white}{\protect\rule[0.5ex]{0.2em}{1pt}}\textcolor{gray}{\protect\rule[0.5ex]{0.2em}{1pt}}\textcolor{white}{\protect\rule[0.5ex]{0.2em}{1pt}}}
 \newcommand{\graydashdotted}{\textcolor{gray}{\protect\rule[0.5ex]{0.2em}{1pt}}\textcolor{white}{\protect\rule[0.5ex]{0.2em}{1pt}}\textcolor{gray}{\protect\rule[0.5ex]{0.4em}{1pt}}\textcolor{white}{\protect\rule[0.5ex]{0.2em}{1pt}}\textcolor{gray}{\protect\rule[0.5ex]{0.2em}{1pt}}}
\newcommand{\blueline}{\textcolor{Blue}{\protect\rule[0.5ex]{1.2em}{1pt}}}
\newcommand{\bluedashed}{\textcolor{Blue}{\protect\rule[0.5ex]{0.2em}{1pt}}\textcolor{white}{\protect\rule[0.5ex]{0.2em}{1pt}}\textcolor{Blue}{\protect\rule[0.5ex]{0.2em}{1pt}}\textcolor{white}{\protect\rule[0.5ex]{0.2em}{1pt}}\textcolor{Blue}{\protect\rule[0.5ex]{0.2em}{1pt}}\textcolor{white}{\protect\rule[0.5ex]{0.2em}{1pt}}}
\newcommand{\bluedashdotted}{\textcolor{Blue}{\protect\rule[0.5ex]{0.1em}{1pt}}\textcolor{white}{\protect\rule[0.5ex]{0.2em}{1pt}}\textcolor{Blue}{\protect\rule[0.5ex]{0.4em}{1pt}}\textcolor{white}{\protect\rule[0.5ex]{0.2em}{1pt}}\textcolor{Blue}{\protect\rule[0.5ex]{0.1em}{1pt}}}
\newcommand{\bluedotted}{\textcolor{Blue}{\protect\rule[0.5ex]{0.1em}{1pt}}\textcolor{white}{\protect\rule[0.5ex]{0.2em}{1pt}}\textcolor{Blue}{\protect\rule[0.5ex]{0.1em}{1pt}}\textcolor{white}{\protect\rule[0.5ex]{0.2em}{1pt}}\textcolor{Blue}{\protect\rule[0.5ex]{0.1em}{1pt}}}

\DeclareRobustCommand{\graystar}{\raisebox{0.5pt}{\tikz{\draw[gray,solid,line width = 1.0pt,fill=white](0.0mm,0.8mm)--++(0.0mm,0.8mm)++(0.0mm,-0.8mm)--++(0.6472mm,0.4702mm)++(-0.6472mm,-0.4702mm)--++(0.4702mm,-0.6472mm)++(-0.4702mm,0.6472mm)--++(-0.4702mm,-0.6472mm)++(0.4702mm,0.6472mm)--++(-0.6472mm,0.4702mm);}}}

\makeindex             

\begin{document}

\title*{Direct numerical simulation of turbulent open channel flow: Streamwise turbulence intensity scaling and its relation to large-scale coherent motions}
\titlerunning{Direct numerical simulation of turbulent open channel flow}
\author{Christian Bauer$^1$, Yoshiyuki Sakai$^2$ and Markus Uhlmann$^3$}
\authorrunning{Bauer et al.} 
\institute{$^1$Institute of Aerodynamics and Flow Technology, German Aerospace Center, Germany, \email{christian.bauer@dlr.de}\\
$^2$  TUM School of Engineering and Design, Technical University of Munich, Germany \\
$^3$ Institute for Hydromechanics, Karlsruhe Institute of Technology, Germany 
}

%
%
\maketitle

\abstract{
We conducted direct numerical simulations of turbulent open channel flow (OCF) and closed channel flow (CCF) of friction Reynolds numbers up to $\mathrm{Re}_\tau \approx 900$ in large computational domains up to $L_x\times L_z=12\pi h \times 4\pi h$ to analyse the Reynolds number scaling of turbulence intensities.
Unlike CCF, our data suggests that the streamwise turbulence intensity in OCF scales with the bulk velocity for $\mathrm{Re}_\tau \gtrsim 400$.
The additional streamwise kinetic energy in OCF with respect to CCF is provided by larger and more intense very-large-scale motions in the former type of flow.
Therefore, compared to CCF, larger computational domains of $L_x\times L_z=12\pi h\times 4\pi h$ are required to faithfully capture very-large-scale motions in OCF---and observe the reported scaling.
OCF and CCF turbulence statistics data sets are available at \url{https://doi.org/10.4121/88678f02-2a34-4452-8534-6361fc34d06b} 
}
\vspace{-1ex}
\section{Introduction}
\label{sec:intro}
\vspace{-1ex}
Plane Poiseuille flow, also known as closed channel flow (CCF), is one of the most studied canonical flows by means of direct numerical simulations (DNSs,~\cite{Kim1987,Moser1999,Hoyas2006,Lee2015,Oberlack2022}). 
The numerical domain is defined by doubly-periodic boundary condition in the stream- and spanwise directions, and impermeable no-slip walls at the bottom and the top.
Conversely, less attention has been paid to open channel flow (OCF), where one of the no-slip walls is replaced by a free-slip plane, despite its even more direct relevance in environmental flows.
Most of the early OCF DNSs were restricted to small friction Reynolds numbers $Re_\tau = u_\tau h/\nu$, based on friction velocity $u_\tau$, channel height $h$ and  kinematic viscosity $\nu$, and/or small computational domains.
Recently, Yao et al.~ \cite{Yao2022} compared CCF and OCF by means of DNS data in computational domains of $L_x \times L_z=8\pi h \times 4 \pi h$ for Reynolds numbers up to $Re_\tau = 2000$. 
In agreement with experiments~\cite{Duan2020,Duan2021} and DNSs~\cite{Bauer2015}, they reported that so-called very-large-scale motions (VLSMs)---which are a common feature of wall-bounded turbulent flows---are more energetic and larger in OCF than in CCF.
Generally, VLSMs become more energetic with increasing Reynolds number~\cite{Bauer2015,Duan2021,Yao2022}, leading to the failure of wall scaling of the streamwise intensity in CCF~\cite{Hoyas2006}.
%
While for CCF the streamwise turbulence intensity scales neither with the friction velocity nor with the bulk velocity, Afzal et al.~\cite{Afzal2009} suggested that the streamwise turbulence intensity in OCF scales with the bulk velocity based on an experimental data set.
Other experimental studies, such as Duan et al.~\cite{Duan2020,Duan2021}, on the contrary, show the---relatively scattered---streamwise turbulence intensity scaled in wall units together with universal scaling functions proposed by Nezu and Rodi~\cite{Nezu1986}.
Thus, additional highly accurate OCF data sets are required to answer the question whether the streamwise turbulence intensity in OCF scales in bulk units.\par
%
Since the streamwise length of VLSMs is an order of magnitude larger than the channel half width, exceptionally large domains are required to faithfully capture VLSMs in DNS~\cite{Lozano2014,Feldmann2018}.
Moreover, Pinelli et al.~\cite{Pinelli2022} found that the small scales---responsible for the mass transfer across the free surface in OCF---are spatially organised by VLSMs,
which requires an additional grid refinement towards the free surface.
%
In the present study, we investigate the scaling of the streamwise turbulence intensity in OCF by means of DNSs.
For this purpose, we generate new OCF and CCF DNS datasets at comparable Reynolds numbers up to  $\mathrm{Re}_\tau=900$ in identical computational domains of $L_x \times L_z = 12\pi h \times 4\pi h$, with grid refinements towards no-slip and free-slip walls.
By removing Reynolds number or domain size discrepancies between OCF and CCF, we are able to quantify the impact of the free-slip boundary condition on turbulence one-point statistics in the cleanest possible manner.\par
%
\vspace{-1ex}
\section{Methodology}
\label{sec:mehodology}
\vspace{-1ex}
We solve the incompressible Navier-Stokes equations in their wall-normal velocity/vorticity formulation with a pseudo-spectral representation of the flow variables~\cite{Kim1987}. 
We apply periodic boundary conditions in the stream- and the spanwise directions, the smooth no-slip wall at the bottom and the free-slip boundary condition at the top of the channel. 
Although the numerical method has shown its validity in numerous CCF simulations~\cite{Kim1987,Moser1999,DelAlamo2003}, the introduction of a free-slip surface has to be regarded separately. A Chebyshev-Gauss-Lobatto grid, which is adopted in the surface-normal direction provides a grid refinement both towards the wall and the free surface.
We set up six OCF and four CCF cases, where the Reynolds number and the computational domain length are varied (cf.~table \ref{tab:cases}). 
\begin{table} 
\centering
\caption{Turbulent channel flow simulation cases. $N_x$, $N_y$ and $N_z$ are the number of grid points in the streamwise, wall-normal and spanwise direction; $\Delta x^+$, the streamwise grid spacing; $\Delta z^+$, the spanwise grid spacing; $\Delta y^+_{max}$, the maximum grid spacing in the wall-normal direction. $\Delta T$ is the temporal averaging interval for turbulent statistics. Prefix ``O'' indicates OCF cases, ``C'' CCF cases.}
\begin{tabularx}{1.00\linewidth}{Xrrrrrrrrrr}
\hline
Case                      & O200      & O400            & O600 & O900L4 & O900L8 & O900L12  & C200 & C400 & C600 & C900        \\
\hline
$Re_\tau$                 & \num{200} & \num{399}       & \num{596}  &\num{899}&\num{895}&\num{894}    &\num{200}&\num{397}&\num{593}& \num{890}      \\
$Re_b$                    & \num{3170}& \num{6969}      & \num{11047}&\num{17512}&\num{17512}& \num{17512}              &\num{3170}&\num{6969}&\num{11047}& \num{17512}       \\
$L_x/h$                   & $12\pi$   & $12\pi$         & $12\pi$    & $4\pi$ & $8\pi$ & $12\pi$      & $12\pi$&$12\pi$ & $12\pi$ & $12\pi$      \\
$L_z/h$                   & $4\pi$    & $4\pi$          & $4\pi$     & $2\pi$ & $4\pi$ & $4\pi$       & $4\pi$ & $4\pi$ & $4\pi$ &$4\pi$        \\
$N_{x}$                   & \num{768} & \num{1536}      &\num{1536}  &\num{1536}&\num{2048}&\num{3072}    &\num{768}&\num{1536}&\num{2048}&\num{3072}       \\
$N_{y}$                   & \num{129} & \num{193}       &\num{257}   &\num{385}&\num{385}&\num{385}       &\num{129} & \num{193}       &\num{257}  &\num{385}       \\
$N_{z}$                   & \num{512} & \num{1024}      &\num{1536}  &\num{1024}&\num{2048}&\num{2048}   & \num{512} & \num{1024}      &\num{1536} &\num{2048}       \\
$\Delta x^+$              & \num{9.8} & \num{9.8}       & \num{14.6}  &\num{7.4}&\num{11.0}&\num{11.0}                 &\num{9.8}&\num{9.8}&\num{14.6}&\num{11.0}       \\
$\Delta z^+$              & \num{4.9} & \num{4.9}       &\num{4.8}    &\num{5.5}&\num{5.5}&\num{5.5}                  &\num{4.9}&\num{4.9}&\num{4.8}&\num{5.5}       \\
$\Delta y^+_{max}$        & \num{2.5}  & \num{3.3}      & \num{3.7}   &\num{3.7}&\num{3.7}&\num{3.7}                  &\num{4.9}&\num{6.5}&\num{7.3}&\num{7.3}       \\
        $\Delta T u_b/h$          & \num{8660} &  \num{1925}    &  \num{1460} &\num{417}&\num{570}&  \num{1054}             &\num{8600}&\num{3260}&\num{1757}&\num{1013}       \\
        $\Delta T u_\tau^2/\nu$   & \num{109700} & \num{43900}  & \num{46980} &\num{19080}&\num{26100}&  \num{44700}           &\num{108200}&\num{73730}&\num{55960}&\num{45820}       \\
\end{tabularx}
\label{tab:cases}
\end{table}
Hereinafter, the velocity fluctuation in $i$-direction is defined as $u^\prime_i = u_i - \langle u_i \rangle$, where angular brackets denote averaging over both homogeneous directions and time. Normalisation in viscous or ``wall units'' is indicated by the $+$ superscript. Corresponding scales are $u_\tau$ and the viscous length scale $\delta_\nu=\nu/u_\tau$. Characteristic scales in ``bulk units'', on the other hand, are the bulk velocity $u_b$ and the channel height $h$, which forms the bulk Reynolds number $Re_b = u_b h / \nu$.

\vspace{-1ex}
\section{Computational domain size}
\label{sec:domain}
\begin{figure}[t]
\centering
\includegraphics[width=\linewidth]{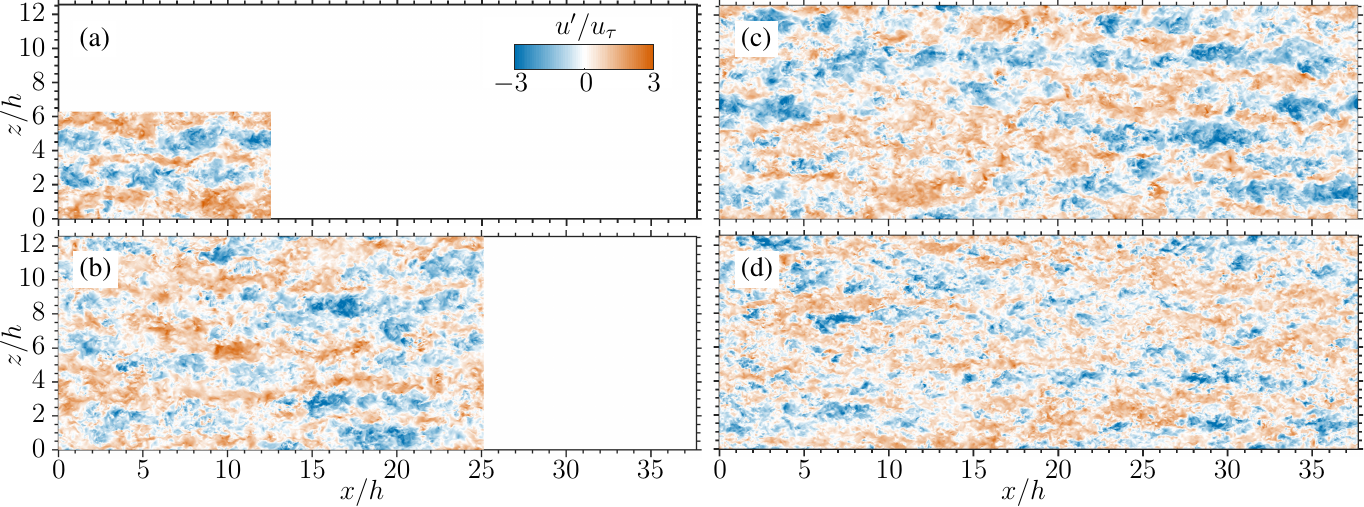}
\caption{Instantaneous streamwise velocity fluctuations $u'^+$ in a wall-parallel plane directly below the free surface in OCF (a,b,c) and the vicinity of the channel centreline in CCF (d, $y/h \approx 0.95$). (a) O900L4; (b) O900L8; (c) O900L12; (d) C900.}
\label{fig:up_compare}
\end{figure}

\begin{figure}[b] 
\sidecaption
\centering
\includegraphics[width=0.64\linewidth]{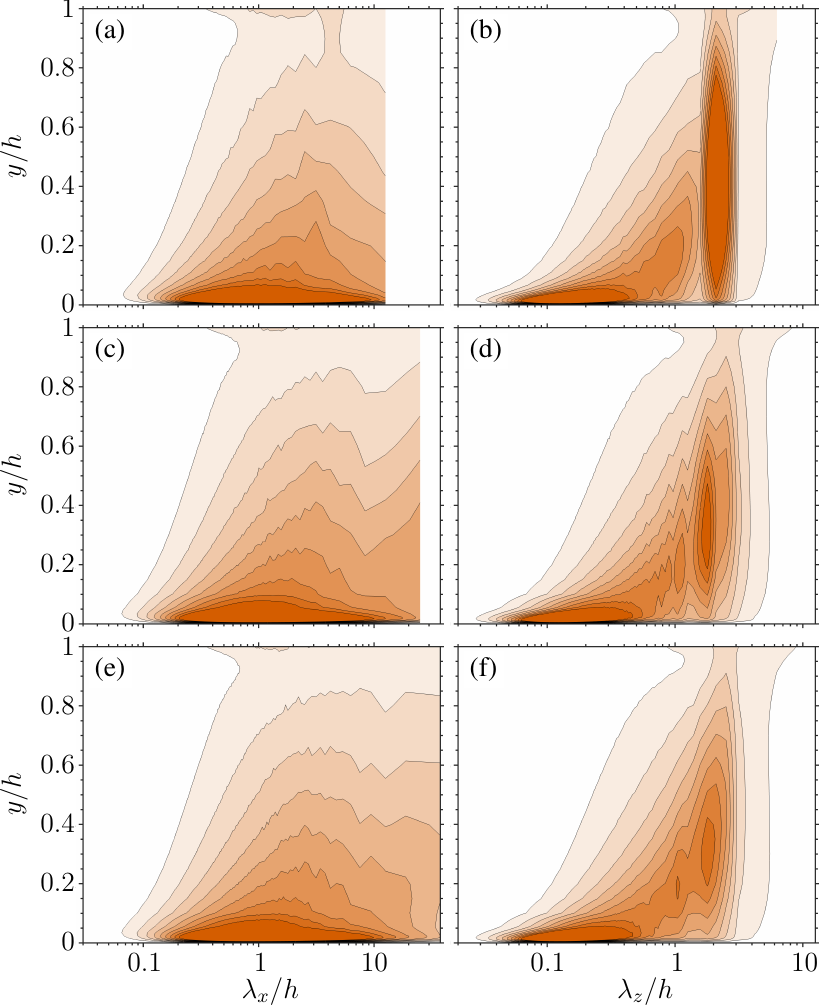}
\caption{Contour lines of the one-dimensional pre-multiplied energy spectra of the streamwise velocity fluctuations as a function of the wavelength and the wall distance. (a,c,e) Streamwise spectra $\kappa_x\phi^+_{uu}(\lambda_x)$; (b,d,f) spanwise spectra $\kappa_z\phi^+_{uu}(\lambda_z)$. (a,b) O900L4; (c,d) O900L8; (e,f) O900L12. Spectra are normalised by their peak values. Values range from 0 (white) to 0.5 (orange) with an increment of 0.05. }
\label{fig:pmsy}
\end{figure}

\vspace{-1ex}
In the following, the effect of the constraints upon large-scale coherent motions in the vicinity of the free surface due to the finite size of the computational domain is investigated.
Figure~\ref{fig:up_compare} displays instantaneous realisations of the streamwise fluctuating velocity for the three open channel simulation cases with different computational domain size O900L4, O900L8, and O900L12 in the vicinity of the free-slip boundary ($y/h\approx0.95$), together with closed channel data for comparison.
A coherent structure with a streamwise extent comparable to the domain length  $l_x \approx 12\pi h$ and a spanwise extent of  $l_z \approx 2h$ is clearly visible in figure~\ref{fig:up_compare}(c) for O900L12 at $z/h\approx 9$, but not for the smaller boxes (a,b), where the flow domain interferes with the natural extensions of such a structure.
Moreover, the structures in the vicinity of the free-slip boundary in OCF appear to be more intense and larger than the structures near the centreline of CCF (figure~\ref{fig:up_compare}d).
Thus, larger computational domains than in CCF are required to fully capture these scales.
Figure~\ref{fig:pmsy} displays one-dimensional pre-multiplied energy spectra of the streamwise velocity fluctuations for $\mathrm{Re}_\tau\approx900$ and different computational domain length as a function of the wavelength and the wall distance.
Although at first glance both the streamwise and the spanwise pre-multiplied energy spectra decay at large wavelengths for O900L4 (figure~\ref{fig:pmsy}a,b), in fact larger box sizes are required to physically capture the largest scales that only appear in the larger domains.
This can be seen in the streamwise spectra for O900L8 (figure~\ref{fig:pmsy}c), where a spectral pile-up at large wavelength occurs.
For O900L12 (figure~\ref{fig:pmsy}e,f), on the other hand, both the streamwise and the spanwise spectra decay at large wavelengths again.
Thus, a domain size of $L_x \times L_z =12\pi h \times 4\pi h$ appears to be sufficiently large to capture a relevant amount of VLSMs.
Moreover, a comparison of the spanwise spectra (figure~\ref{fig:pmsy}b,d,f) indicates an artificially large accumulation of energy at spanwise wavelengths around $\lambda_z/h\approx 2$ in the smallest box ($L_x=4\pi h$, $L_z=2\pi h$).
In summary, as in CCF, VLSMs appear in the bulk flow, penetrating into the buffer layer and therefore through these motions the whole flow domain is influenced by the free-surface layer.
Due to their large spatial extent, larger computational domains than in CCF are needed to fully capture these scales. A domain size of $L_x=12\pi h$ was found to be sufficient at $\mathrm{Re}_\tau=900$.\par
\par

\vspace{-1ex}
\section{Turbulence intensity scaling}
\label{sec:results}
\begin{figure} 
\centering
\includegraphics[]{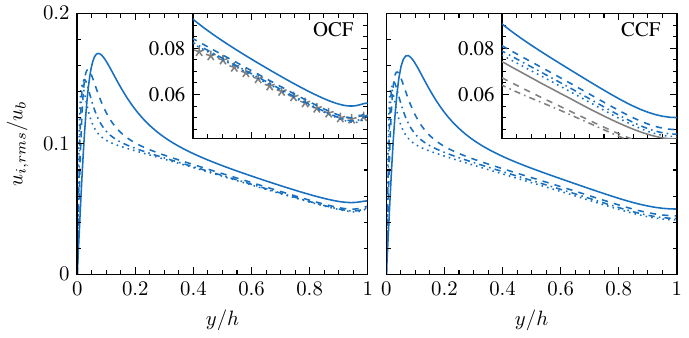}
 \caption{Turbulence intensities normalised by $u_b$ as function of the distance from the wall $y/h$ for OCF (left) and CCF (right): \blueline, $\mathrm{Re}_\tau \approx 200$; \bluedashed, $\mathrm{Re}_\tau \approx 400$; \bluedashdotted, $\mathrm{Re}_\tau \approx 600$; \bluedotted, $\mathrm{Re}_\tau \approx 900$. The insets show a zoom for the upper channel half.
Left: The symbols (\graystar) indicate a profile from OCF measurements at $\mathrm{Re}_\tau=2407$~\cite{Duan2020}.
Right: The gray lines indicate CCF DNS data at $\mathrm{Re}_\tau=2003$ (\grayline, \cite{Hoyas2006}), $\mathrm{Re}_\tau=5186$ (\graydashed, \cite{Lee2015}), $\mathrm{Re}_\tau=10049$ (\graydashdotted, \cite{Oberlack2022}).}
\label{fig:urms}
\end{figure}

\vspace{-1ex}
Let us now focus on the variation with Reynolds number of the turbulence intensities normalised with the bulk velocity $u_b$.
Both open and closed channel turbulence intensities obtained from DNSs with the largest computational domain ($L_x \times L_z=12\pi h  \times 4 \pi h$) are presented in figure~\ref{fig:urms}.
While the spanwise and wall-normal contributions to the turbulence intensity scale well with the friction velocity $u_\tau$ in the bulk region of the flow for $Re_\tau \ge 400$  for both flow configurations (figure omitted), the streamwise contribution differs.
As figure \ref{fig:urms}(a) indicates, the streamwise turbulence intensity in OCF appears to scale with the bulk velocity $u_b$ for $Re_\tau \ge 400$ and $y/h \gtrsim 0.3$, whereas in the closed channel case it neither scales with $u_\tau$ (figure omitted) nor with $u_b$ (figure \ref{fig:urms}b).
This scaling property of OCF is in qualitative agreement with the experimental study of OCF by \cite{Afzal2009}, who observed the scaling of the outer flow streamwise turbulence intensity with $u_b$ for $1260 \le Re_\tau \le 5280$.
Besides, the streamwise turbulence intensity profile obtained from recent OCF measurements by \cite{Duan2020,Duan2021} at $\mathrm{Re}_\tau=2407$ normalised with $u_b$ collapses well with our data (figure~\ref{fig:urms}a inset).
The above observation differs significantly from CCF, where the spectral signature of VLSMs first appears at $\mathrm{Re}_\tau\approx 600$ (figure omitted) and the contribution of VLSMs to the streamwise turbulence intensity is much lower than in OCF~\cite{Duan2020}. Even at $\mathrm{Re}_\tau = 10049$, which is currently the highest Reynolds number ever been achieved by a CCF DNS~\cite{Oberlack2022}, the bulk scaling of the streamwise turbulence intensity has not been observed. Hence, it is remarkable that OCF streamwise turbulence intensity scales with $u_b$ at a Reynolds number as low as $\mathrm{Re}_\tau=400$, if a sufficiently large numerical domain is used. \par

\vspace{-1ex}
\section{Conclusion}
\label{sec:conclusion}
\vspace{-1ex}
We performed DNSs of OCF and CCF in computational domains up to $L_x\times L_z=12\pi h \times 4\pi h$ and Reynolds numbers up to  $\mathrm{Re}_\tau=900$.
The enhancement of VLSMs in OCF leads to more stringent requirements on the domain size than in the closed channel counterpart.
Since VLSMs appear at lower Re in OCF, a domain size $L_x \ge 12\pi h$ is already required at the moderate value $Re_\tau=400$.
Unlike CCF, the streamwise turbulence intensity in OCF scales with the bulk velocity $u_b$ for $Re_\tau \ge 400$ and $y/h \gtrsim 0.3$, which is attributed to the larger and more intense VLSMs in OCF.
Although it was suggested by the experiment of Afzal et al.~\cite{Afzal2009}, this is the first time the bulk scaling of the streamwise turbulence intensity has been observed in DNS data.
In addition, only with $L_x = 12\pi h$ the bulk-scaling profile appears in the $Re_\tau=400$ case (figure omitted), which indicates that VLSMs in OCF already play a significant role at this Reynolds number.
Thus, it is of critical importance to use sufficiently large numerical domains.
Finally, both OCF and CCF turbulence statistics data sets are available online~\cite{BauerData2023}.\par

 \begin{acknowledgement}
Many fruitful discussions with Genta Kawahara throughout this work are gratefully acknowledged.  The simulations were carried out at UC2 of SCC Karlsruhe and at DLR's CARA cluster. The computer resources and assistance provided by these centres are gratefully acknowledged.
 YS and MU acknowledge funding by DFG through grant UH~242/3-1.
 \end{acknowledgement}
 \vspace{-5ex}

 

\end{document}